\documentclass[12pt]{article}
\usepackage{graphicx}
\usepackage{lineno}

\newcommand\pubnumber{}
\newcommand\pubdate{\today}

\def\institute{Laborat\'{o}rio de Instrumenta\c{c}\~{a}o e F\'{i}sica Experimental de Part\'{i}culas, Braga, Portugal}
\def\support{\footnote{The author is funded by the grant LIP/50007 through FCT, COMPETE2020-Portugal2020, FEDER, POCI-01-0145-FEDER-007334 and by the FCT project CERN/FIS-NUC/0005/2015.}}

\def\Title#1{\begin{center} {\Large #1 } \end{center}}
\def\Author#1{\begin{center}{ \sc #1} \end{center}}
\def\Address#1{\begin{center}{ \it #1} \end{center}}

\newcommand\pubblock{\rightline{\begin{tabular}{l} \pubnumber\\
         \pubdate  \end{tabular}}}
\newenvironment{Abstract}{\begin{quotation}  }{\end{quotation}}
\newenvironment{Presented}{\begin{quotation} \begin{center} 
             PRESENTED AT\end{center}\bigskip 
      \begin{center}\begin{large}}{\end{large}\end{center} \end{quotation}}

\def\beq{\begin{equation}}
\def\eeq#1{\label{#1}\end{equation}}
\def\eeqn{\end{equation}}

\def\beqa{\begin{eqnarray}}
\def\eeqa#1{\label{#1}\end{eqnarray}}
\def\eeqan{\end{eqnarray}}

\let\bar=\overbar

\def\Dslash{\not{\hbox{\kern-4pt $D$}}}
\def\dslash{\not{\hbox{\kern-2pt $\del$}}}

\def\msb{{\bar{\ssstyle M \kern -1pt S}}}

\def\met{E_{\rm T}^{\rm miss}}

\begin{document}
\begin{titlepage}
\pubblock

\vfill
\Title{Overview of the vector-like quark searches with the LHC data collected by the ATLAS detector}
\vfill
\Author{ J. P. Araque\support\\On behalf of the ATLAS Collaboration}
\Address{\institute}
\vfill
\begin{Abstract}
	In 2012 the discovery of the Higgs boson by the ATLAS and CMS collaborations set a milestone in particle physics by finding the missing piece of the Standard Model. Nonetheless some questions are still open: the origin of the mass of the neutrino and finding the missing candidate for dark matter are some examples. One of the main issues with the Standard Model is the hierarchy problem which appears when trying to go to high energy scales at which the theory cannot accommodate corrections large enough to explain the observed nature. Vector-like quarks appear naturally in some non-supersymmetric models which try to find a solution for this issue. The different searches for vector-like quarks within the ATLAS Collaboration using data collected at a centre-of-mass energy of 8 and 13~TeV with integrated luminosities of $20.3\rm{~fb^{-1}}$ (8~TeV) and $3.2\rm{~fb^{-1}}$ and $14.7\rm{~fb^{-1}}$ (13~TeV) are discussed. 
	\end{Abstract}
\vfill
\begin{Presented}
$9^{th}$ International Workshop on Top Quark Physics\\
Olomouc, Czech Republic,  September 19--23, 2016
\end{Presented}
\vfill
\end{titlepage}
\def\thefootnote{\fnsymbol{footnote}}
\setcounter{footnote}{0}

\section{Introduction}

Vector-like quarks (VLQs) naturally appear in several non-supersymmetric theories which propose a solution to the hierarchy problem such as extra-dimension theories or composite and little-Higgs models. The ATLAS Collaboration~\cite{ATLAS} designed a comprehensive strategy to search for these new heavy particles which can be found in different multiplets (singlets, doublets and triples) focusing on final states which are highly sensitive to the three possible decay channels: $Wq$, $Zq$ and $Hq$, where $q$ is a third generation Standard Model (SM) quark. Vector-like quarks can be produced in pairs, in which case the production takes place through QCD interactions and is rather model independent, and also produced singly, in which case the electro-weak (EW) coupling is directly related to the production rate and therefore the analyses will be sensitive to it. In these proceedings an overview of the Run-1 analyses and the most up-to-date results from Run-2 are presented.

\section{Overview of 8~TeV analyses}

\begin{figure}[!t]
\centering
\includegraphics[width=.38\linewidth]{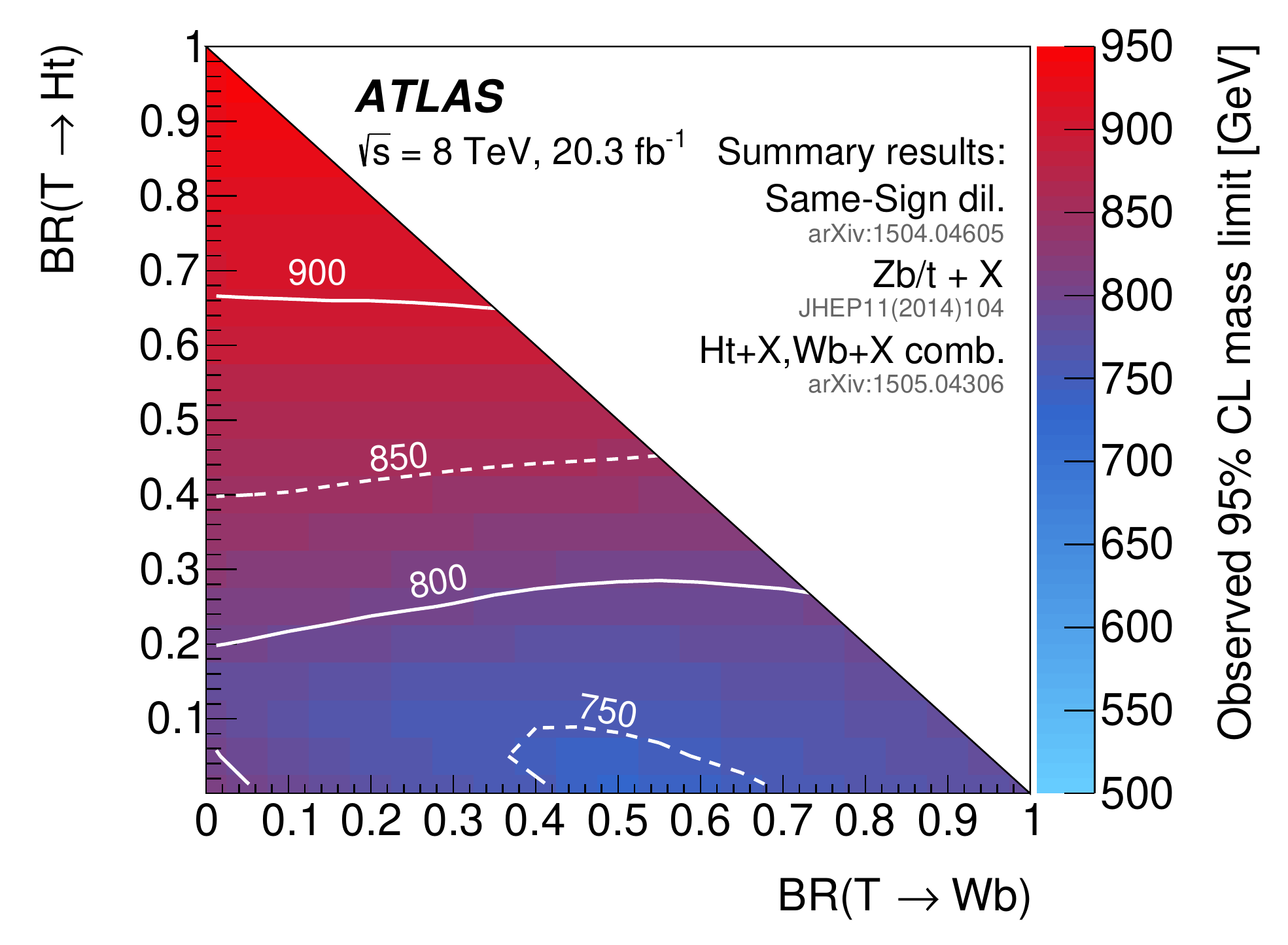}
\includegraphics[width=.38\linewidth]{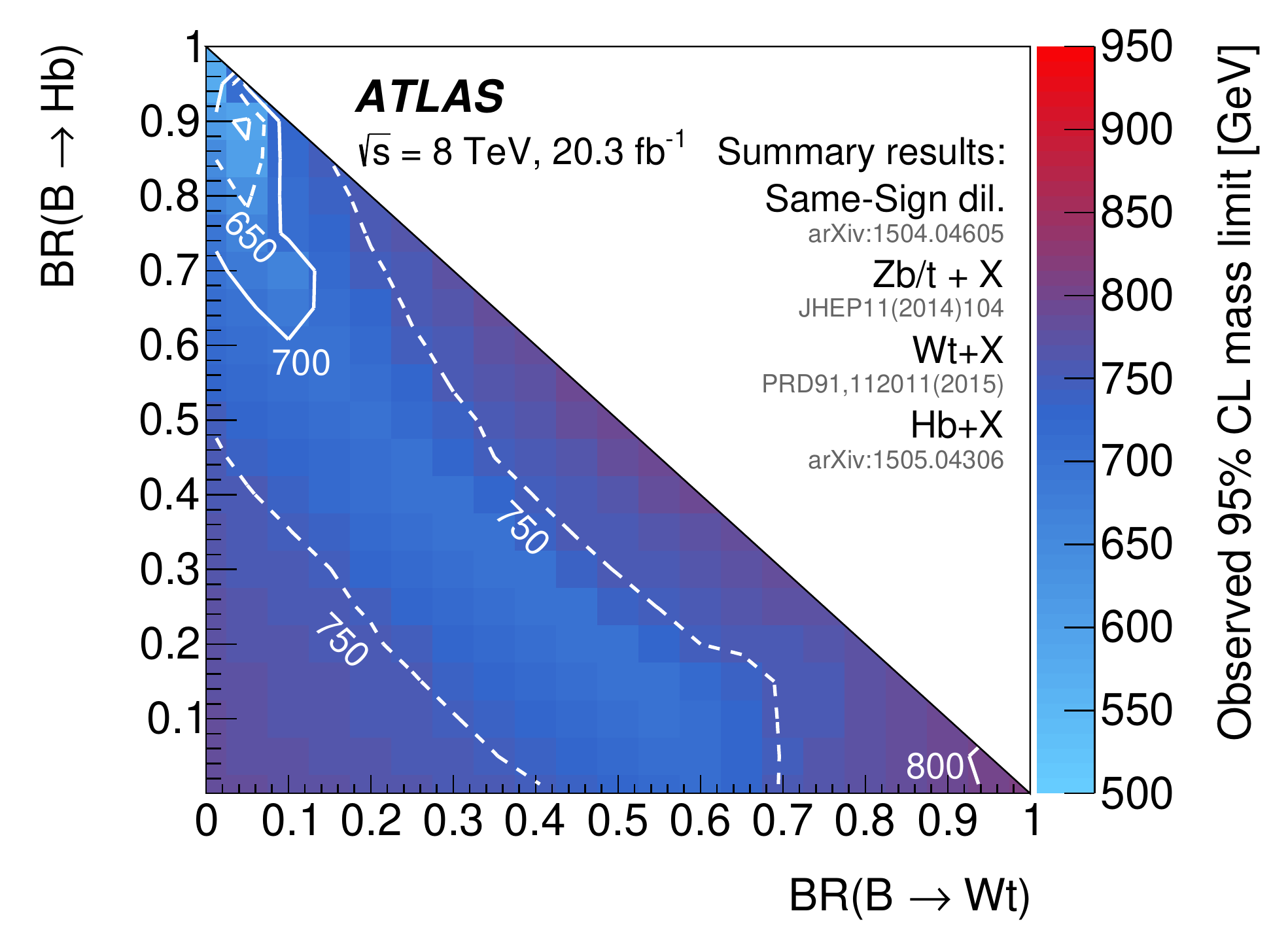}\\
  \includegraphics[width=.31\textwidth]{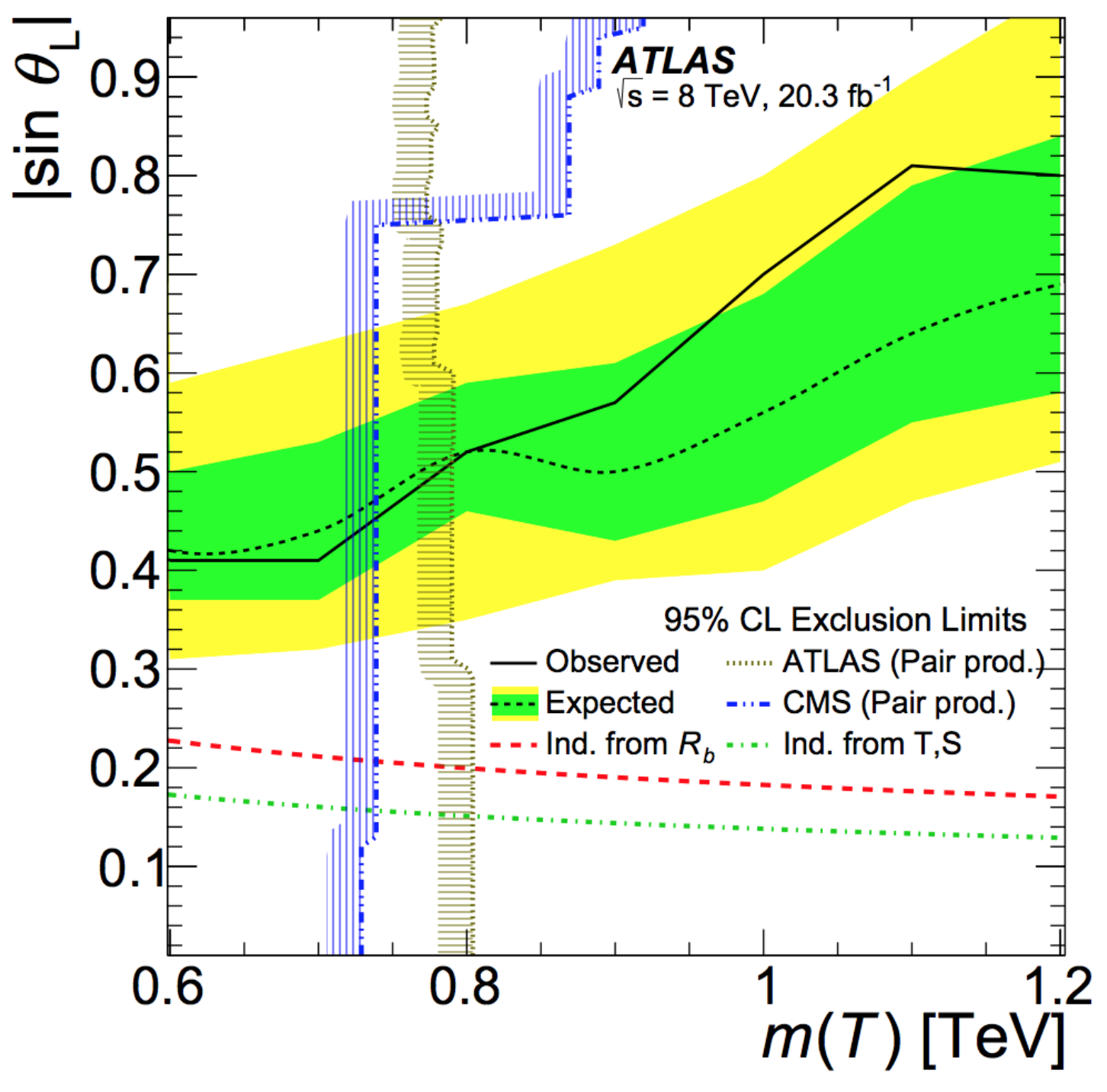}
  \includegraphics[width=.38\linewidth]{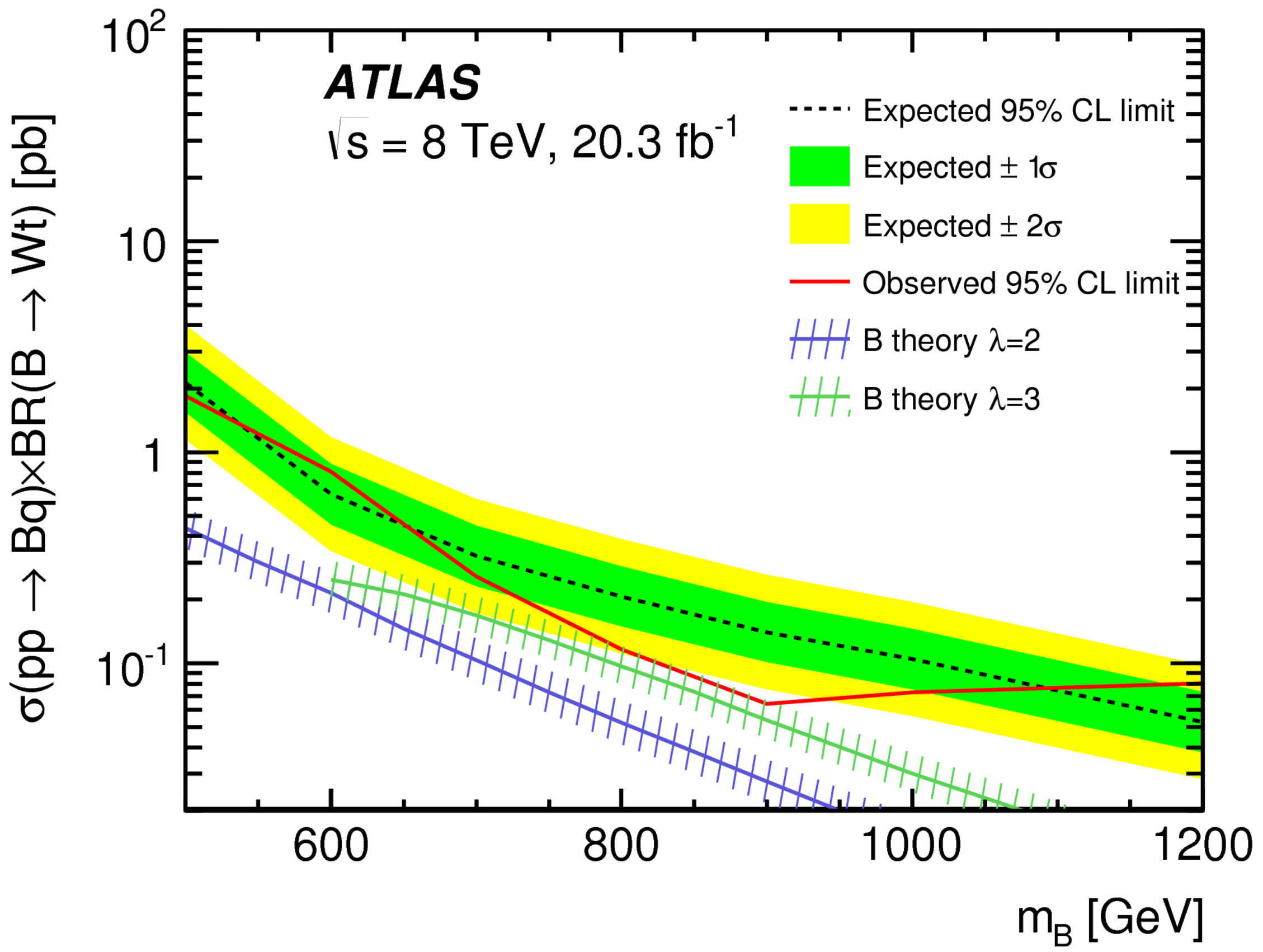}
\caption[Limits exclusions]{Summary of the excluded masses of the VLQs at 8~TeV for the $T$ quark (top-left) and $B$ quark (top-right) as a function of the BRs~\cite{htx8}. Upper exclusion limit for $|\sin\theta|$ as a function of the $T$ mass (bottom-left)~\cite{wbsing8} and upper exclusion limits of the cross-section times BR as a function of the $B$ (bottom-right)~\cite{wtsing8} for single production 8~TeV analyses.}
\label{fig:8tev}
\end{figure}

The strategy followed at 8~TeV (which still holds for the 13~TeV analyses) was to focus on different topologies which are highly sensitive to different decay channels. The analysis here designed as $Zb/t+X$ focused on multilepton final states with at least one pair of opposite-sign same-flavour (OS-SF) pair which is used to reconstruct a $Z$ boson~\cite{ZTAG}. This analysis is highly sensitive to high branching-ratios (BRs) to the $Z$ boson decay channel. The same-sign (SS) analysis~\cite{SS8} focus on final states with a pair of SS leptons and $b$-tagged jets. Depending on the VLQ considered (either a $T$ or $B$ quark with an electric charge of $2/3$ and $-1/3$ respectively) this analysis is more sensitive to the $W$ decay channel (in the case of the $B$ quark) or the Higgs  decay channel (in the case of the $T$ quark). The $Wt+X$ analysis~\cite{wtx8} focuses on the single-lepton plus jets final states with at least one $b$-tagged jet. This makes the analysis sensitive to the $W$ decay channel of the $B$ quark. The $Ht/Wb+X$ analysis~\cite{htx8} is also focused on lepton+jets final state with a different selection strategy based on the number of jets, $b$-tagged jets and their properties. It is sensitive to the $T\rightarrow Wb/Ht$ decays and with small modifications also allows to probe the $B\rightarrow Hb$ channel. Lower limits on the mass of each VLQ are set at a 95\% confidence level (C.L.) and Figure~\ref{fig:8tev}~(top) shows the summary of the best excluded mass among all analysis at 8~TeV, for the VLQ $T$~(left) and $B$~(right).

In the case of single production, two 8~TeV analyses were published. Both analyses are sensitive to the $W$ decay channel, one of them focusing on the $T$ quark~\cite{wbsing8} and another one targeting the $B$ quark~\cite{wtsing8}. Both are focused on lepton+jets final states (with the second one having a di-lepton channel as well) but with different object selection and event reconstruction that allow for different sensitivities. For the analysis sensitive to the $T$ quark decay, Figure~\ref{fig:8tev}~(bottom-left) shows the observed and expected upper limits  at a 95\% C.L. on the mixing angle ($|\sin\theta_L|$) between the VLQ and the third generation quarks it couples with. Figure~\ref{fig:8tev}~(bottom-right) shows the observed and expected upper limits, in the analysis sensitive to the $B$ quark, at a 95\% C.L. of the cross-section times BR as a function of the $B$ mass.

\section{Overview of 13~TeV analyses}

The strategy adopted at 8~TeV is being followed at 13~TeV, for which several analyses have been released with an integrated luminosity ranging from $3.2\rm{~fb^{-1}}$ to $14.7\rm{~fb^{-1}}$.

\begin{figure}[!t]
\centering
\hspace{0.5cm}\includegraphics[width=.35\linewidth]{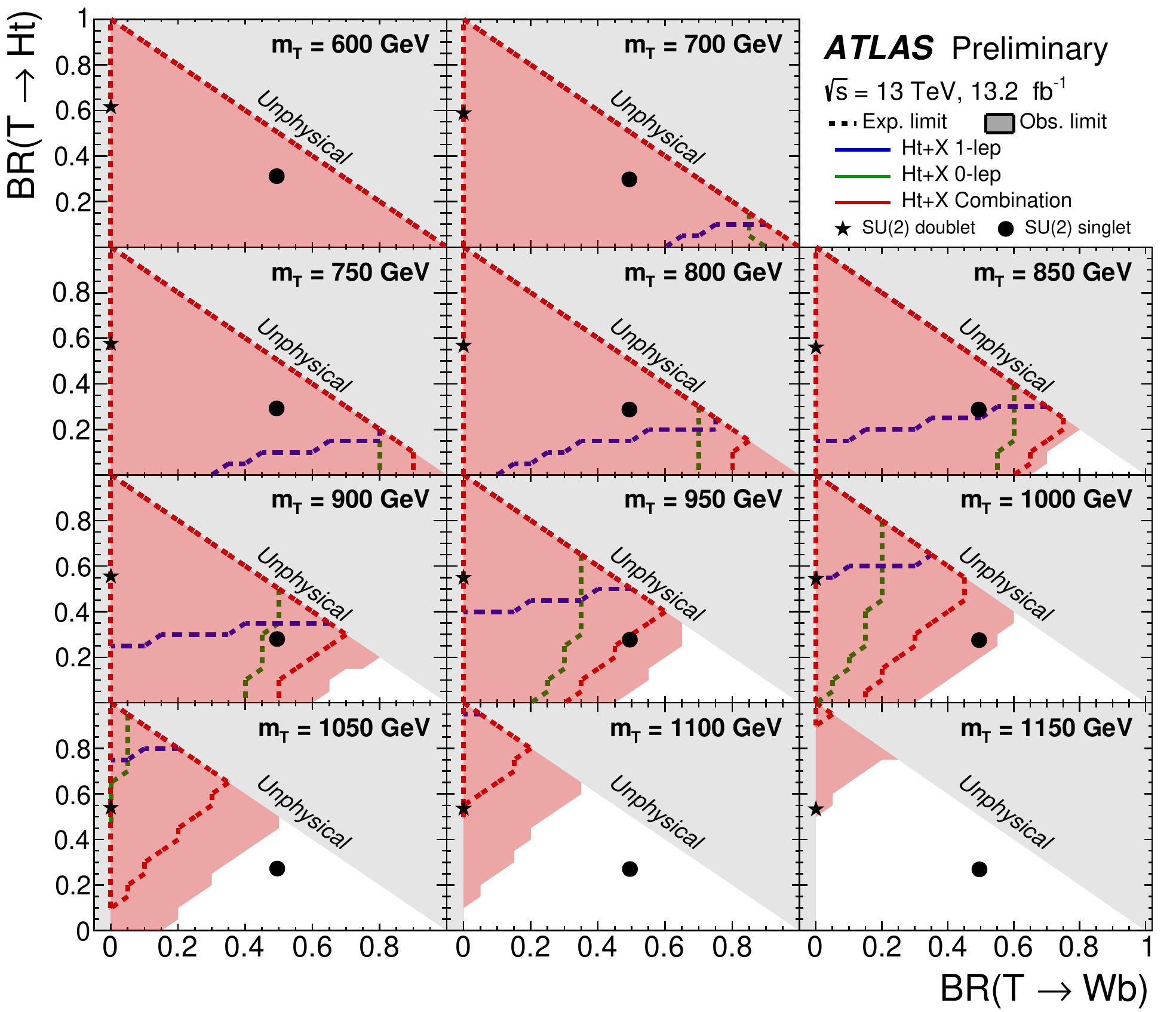}
\hspace{0.8cm}\includegraphics[width=.42\textwidth]{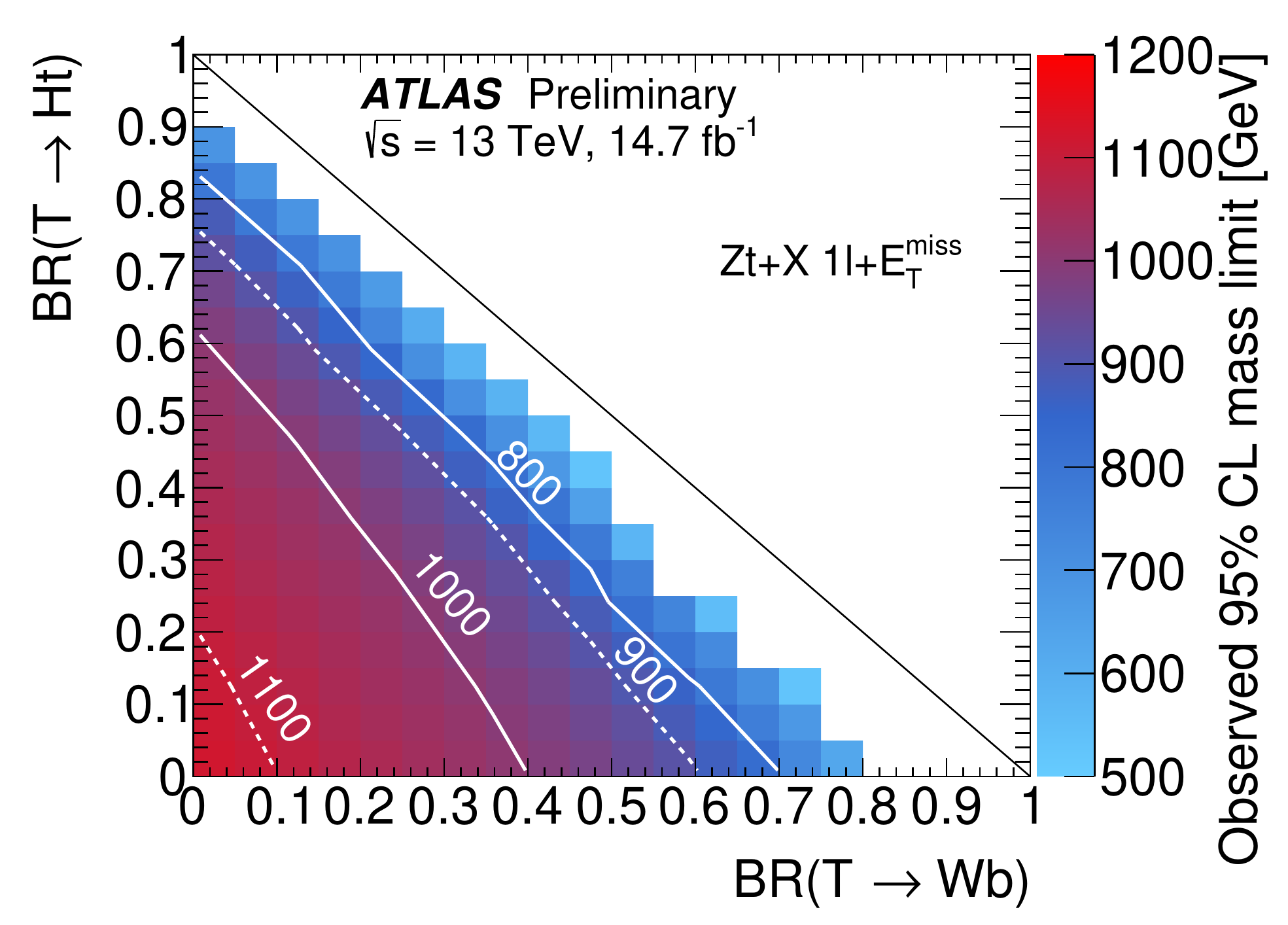}
\includegraphics[width=.42\linewidth]{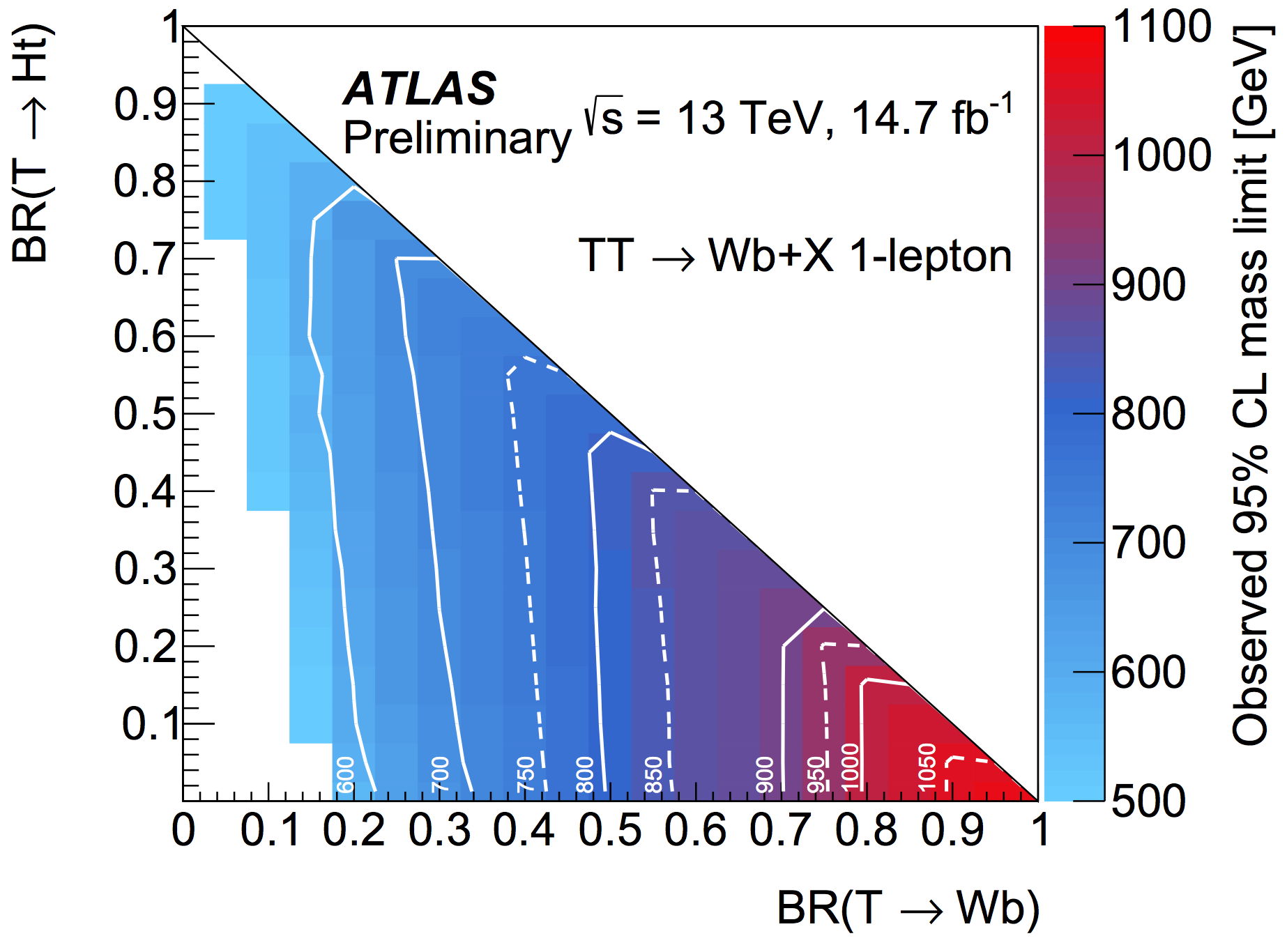}
\includegraphics[width=.42\linewidth]{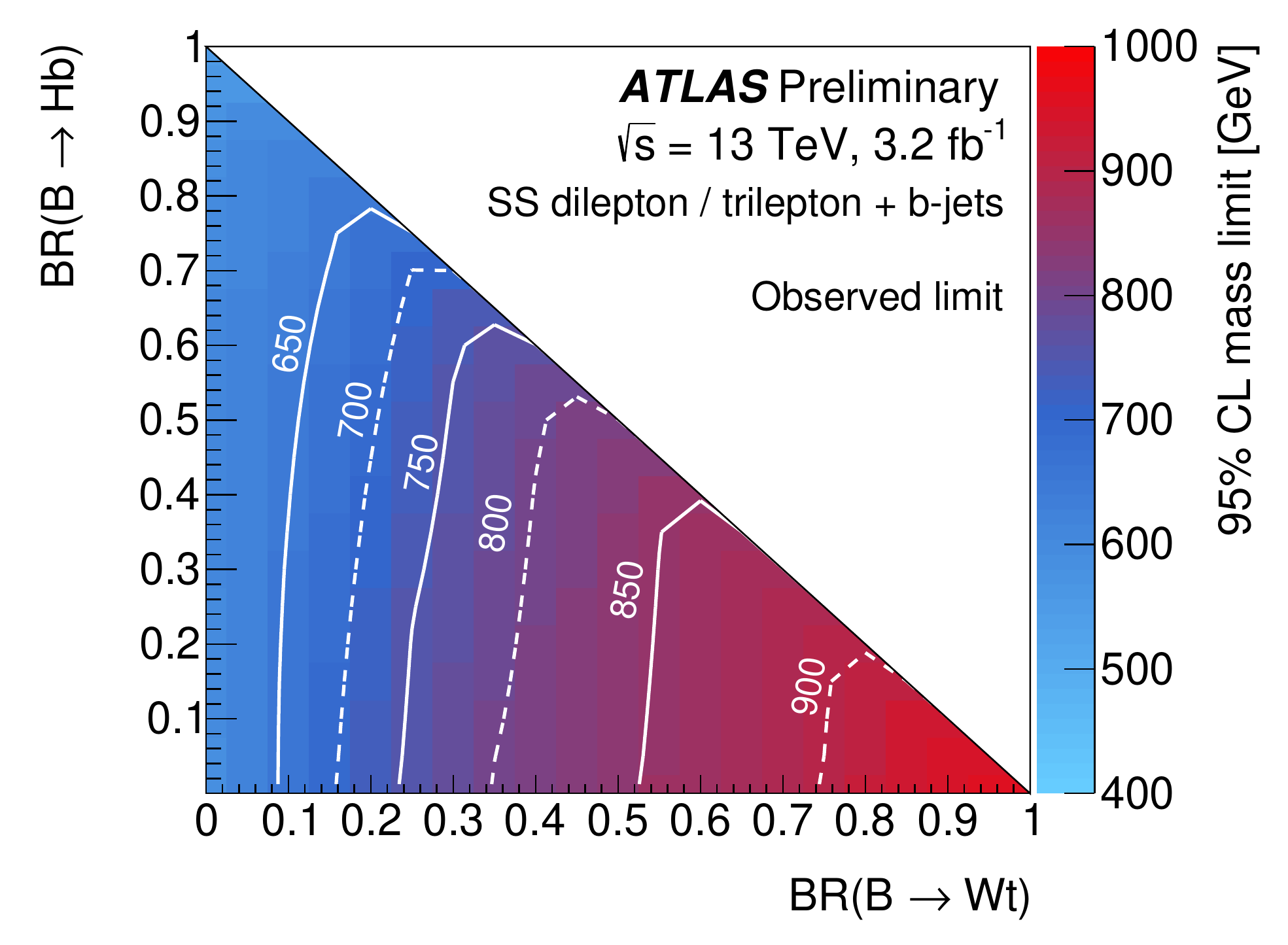}
\caption[Limits exclusions]{Excluded BRs for each mass for the one and zero-lepton channels in the $Ht+X$ analysis at 13~TeV (top-left)~\cite{htx13} and lower mass limit as a function of the BR for the $Zt$ single lepton analysis (top-right)~\cite{ztx13} , $Wb$ analysis (bottom-left)~\cite{wbx13} and SS analysis (bottom-right)~\cite{SS13} at 13~TeV.}
\label{fig:13tev}
\end{figure}

For pair production four analyses have been done targeting different corners of the BR plane. The $Ht+X$ analysis~\cite{htx13} is the update to 13~TeV data, with an integrated luminosity of $13.2\rm~fb^{-1}$, of the $Ht/Wb+X$ analysis published for 8~TeV and focused on single-lepton and zero-lepton final states (to later combine both channels) which includes sensitivity to the $Z$ channel decaying to $\nu\nu$. The analysis defines several regions based on jets and $b$-tagged jets multiplicity which allows to control different background processes. Observed lower limits are set at a 95\% C.L. on the mass of the $T$ quark, excluding masses up to 1.6~TeV. Figure~\ref{fig:13tev}~(top-left) shows the excluded BRs for each mass for the single and zero-lepton channels and the combination, which shows how the inclusion of a zero-lepton channel improves the sensitivity of the analysis. The $Zt+X$ single-lepton analysis~\cite{ztx13} focuses on the $Z$ decay channel of the $T$ quark when the $Z$ boson decays to $\nu\nu$. It requires high $\met$ and jet multiplicity and sets observed lower limits, at a 95\% C.L., on the $T$ mass up-to 1.14~TeV (when BR$(T\rightarrow Zt) = 100$\%). Figure~\ref{fig:13tev}~(top-right) shows the observed lower limit on the mass for each BR. The $Wb+X$ analysis~\cite{wbx13} focus on a lepton+jets final state presenting high sensitivity to the $T\rightarrow Wb$ process. Two orthogonal signal regions are considered (boosted and resolved) which are later combined for the final results. The analysis sets observed lower limits at 95\% C.L. on the  mass of the $T$ quark up to 1.1~TeV (when BR$(T\rightarrow Wb)=100$\%). Figure~\ref{fig:13tev}~(bottom-left) shows the observed lower limit on the mass of the $T$ quark for each BR. The SS analysis~\cite{SS13} is the update of the SS analysis published at 8~TeV. It focuses on a pair of SS leptons with high $\met$ and $b$-jets multiplicity which makes the analysis sensitive to high BR to the $H$ channel for the $T$ and $Y$ quarks and to the $W$ channel for the $B$ quark. Observed lower limits on the mass of the VLQ are set at a 95\% C.L. when a singlet is considered to 0.78~TeV, 0.83~TeV and 0.99~TeV for the $T$, $B$ and $Y$ quarks respectively. Figure~\ref{fig:13tev}~(bottom-right) shows the lower mass limit for each BR for the $T$ quark.

Single production becomes the dominant production mechanism as the higher masses get excluded due to the reduction of the available phase space to produce a pair of VLQs. Although at the beginning of Run-2 pair production is still dominant due to the increase in the centre-of-mass energy, single production analyses are also being developed in ATLAS at 13~TeV. A single production analysis~\cite{wbsing13} has been released with an integrated luminosity of $3.2\rm{~fb}^{-1}$ focused on the $Wb$ decay of the $T$ and $Y$ quarks. This is achieved by focusing on lepton+jets final states with a high $\met$. At the end, observed lower limits are set at a 95\% C.L. on the mass of the VLQ achieving exclusions up to 1.44~TeV for the $Y$ quark. Upper limits are also set on $|\sin\theta|$ which goes as low as ~0.3 and as high as ~0.7 for the $T$ singlet case.

\end{document}